\RequirePackage{etoolbox}
\pretocmd\PackageWarning{%
  \edef\pkgname{#1}\edef\hyperrefname{hyperref}%
  \ifx\pkgname\hyperrefname
    \expandafter\gobblethree
  \fi
}{}{\undefined}
\newcommand*{\gobblethree}[3]{}

\documentclass[a4paper,fleqn,usenatbib]{mnras}

\usepackage[T1]{fontenc}
\usepackage{ae,aecompl}

\usepackage{graphicx}	
\usepackage{amsmath}	
\usepackage{url}
\usepackage{booktabs}
\usepackage{threeparttable}
\usepackage{pdflscape}
\usepackage[table]{xcolor}
\definecolor{lightgray}{gray}{0.9}

\newcommand*\rfrac[2]{{}^{#1}\!/_{#2}}

\title[Statistical analysis of asteroid periods]{Statistical analysis of the ambiguities in the asteroid period
determinations}

\author[M. Butkiewicz-B\k{a}k et al.]{
M. Butkiewicz-B\k{a}k,\thanks{E-mail: mbutek@amu.edu.pl}
T. Kwiatkowski,
P. Bartczak,
G. Dudzi\'{n}ski,
\newauthor
and A. Marciniak
\\
Astronomical Observatory, Faculty of Physics, Adam Mickiewicz University, S{\l}oneczna 36, 60-286 Pozna\'{n}, Poland
}

\date{Accepted 25 May 2017. Received 29 Mar 2017; in original form 25 May 2017}

\pubyear{2017}

\begin{document}
\label{firstpage}
\pagerange{\pageref{firstpage}--\pageref{lastpage}}
\maketitle

\begin{abstract}
Among asteroids there exist ambiguities in their rotation period determinations. They are due to incomplete coverage of the rotation,
noise and/or aliases resulting from gaps between separate lightcurves. To help to remove such uncertainties, basic characteristic of the lightcurves
resulting from constraints imposed by the asteroid shapes and geometries of observations should be identified. 
We simulated light variations of asteroids which shapes were modelled as Gaussian random spheres, with random
orientations of spin vectors and phase angles changed every $5\degr$ from $0\degr$ to $65\degr$.
This produced 1.4 mln lightcurves. For each simulated lightcurve Fourier analysis has been made
and the harmonic of the highest amplitude was recorded.
From the statistical point of view, all lightcurves observed at phase angles $\alpha < 30\degr$, with peak-to-peak amplitudes
$A>0.2$~mag are bimodal. Second most frequently dominating harmonic is the first one, with the 3rd harmonic following right after.
For 1\% of lightcurves with amplitudes $A < 0.1$~mag and phase angles $\alpha < 40\degr$ 4th harmonic dominates.
\end{abstract}




\section{Introduction}
\label{Intro}
One of the basic parameters in physical studies of asteroids is the rotation period.
For a population of bodies undergoing collisional evolution, the distribution of their
spin rates should be close to Maxwellian which is true for larger bodies with diameters
$D > 40$~km \citep{Pra+02}. However, for smaller objects other factors come into play
like the YORP effect, whose influence can be studied by analysing asteroid rotation periods \citep{Ros+09}.
An interesting picture appears when we plot asteroid periods against their diameters.
For objects larger than 150~m there exists a minimum allowable period of 2.2~h while many
of $D < 150$~m asteroids display much shorter periods \citep{Pra+00}.
Those periods also have their limits, although much lower, which are usually atributed
to the effect of rotational fission \citep{Hol+07}. They can be used to test different models
of asteroid interiors.

Determination of most asteroid periods is based on photometric observations which reveal 
brightness variations caused by the rotation of an elongated body of uniform reflectivity.
Albedo variegations on asteroid surfaces are very rare -- more important are craters and other
topografic features which at higher solar phase angles cast shadows adding their contribution to the light variations.

The lack of simple periodicity in the lightcurves can point to non-principal axis rotation.
Such objects usually have long rotation periods and small diameters\citep{Pra+02}.
Their tumbling rotation can result from subcatastrophic impacts \citep{Hen+13}, which do not
lead to the disruption of the coliding bodies.

A synodic period of an asteroid can be derived from its lightcurve by standard methods like 
the phase dispersion minimization \citep{Ste+78} or Fourier series fitting \citep{Har+89}.
A problem appears when results of observations are based on less than full coverage of
a lightcurve and/or contain high level of noise. Also long gaps between individual lightcurves create
an ambiguity in the cycle count which leads to aliases.

\begin{table}
  \caption{Reliability of asteroid periods}
  \label{Tab+code}
  \centering
  \begin{threeparttable}
  \begin{tabular}{lrrrr}
  \toprule
  Selection criteria		&	$U=1$	&	$U=2$	&	$U=3$	&	All	\\
  \midrule
  All data			&	1602	&	11804	&	4000	&	17406	\\
  $D<0.15$~km			&	36	&	114	&	106	&	256	\\
  $D<0.15$~km, $P<2.2$~h	&	17	&	61	&	85	&	163	\\
  \bottomrule
  \end{tabular}
  \begin{tablenotes}[para,flushleft]
  \textbf{Notes.} Number of asteroids with periods of the specified quality code $U$. The last column
  gives the total number of periods for all three classes. Data taken from the LCDB version 3~Feb~2017 \footnote{\url{http://www.minorplanet.info/lightcurvedatabase.html}} \citep{War+09}.
  \end{tablenotes}
  \end{threeparttable}
\end{table}

As the data on asteroids periods became to accumulate, \cite{Har+83} introduced  reliability code $U$
to better distinguish between reliable and uncertain results. $U=1$ is used to indicate a period 
which is based on fragmentary lightcurve and might be completely wrong. $U=2$ is assigned to periods 
where ambiguity exist or results were based on over a half coverage of the rotation. $U=3$ indicates
secure result. Later, this designation system was modified by \cite{War+09} to allow for subclasses,
e.g. $2-$ or $2+$. They represent the subjective assessment of rotation period ('+' indices period with
better and '-' with worse reliabilty than the number alone).

The reliability codes $U$ are used in the most complete database of asteroid rotation periods,
LCDB\footnote{\url{http://www.minorplanet.info/lightcurvedatabase.html}} \citep{War+09}, the last
version of which was compiled on 3~Feb~2017. In Table \ref{Tab+code} we present information on the number
of asteroids included in the LCDB, divided into groups of different reliabilities. Separately we have shown two subgroups:
Very Small Asteroids (VSAs) with diameters smaller than $0.15$~km and Fast Rotating Asteroids (FRAs) selected
from VSAs, whose periods are shorter than the $2.2$~h barrier. Most of VSAs and FRAs belong to near-Earth asteroids
which can only be observed during their close encounters with the Earth, often at high solar phase angles.
Windows of opportunity for those objects are short, and it is more difficult to collect enough photometric data for
reliable period determination.

As one can see, in all groups objects with $U=1$, and especially with $U=2$, present a significant fraction of the total.
This reflects the fact that ambigiuous results can be found in many papers, some of them trying to derive meaningful periods 
from too noisy data -- critical discussion of some of them is presented in \cite{Har+12}.
While the best way to remove period ambiguities is to collect more data, weather patterns, instrumental issues and the like 
can make it impossible. In such cases some a priori knowledge about relation between asteroid shapes and lightcurve morphology
can be useful.

This problem was also discussed by \cite{Cel+89}, who computed light variations of a set of irregularly shaped bodies at zero phase
angle, changing the aspect~(the angle between the viewing direction and the spin axis). They obtained lightcurves with one, two,
three and four maxima and minima per rotation, some of them changing from one type of shape to the other with varying aspect. They pointed out that
such phenomena can lead to incorrect estimate of the rotation period.

Very similar discussion was also provided by \cite{Har+14}, who considered lightcurves produced by simple geometric shapes (with two, three, four and
more sides), observed at low phase angles. They have shown that while typical bimodal light variations are caused by rotating elongated
bodies, triangle or square shaped objects can display low amplitude lightcurves dominated by higher Fourier harmonics. This leads to
ambiguities in the rotation period determinations. They concluded that, for low phase angles, lightcurves with amplitudes of $0.2-0.3$ may be
dominated by other harmonics, making the derived period ambiguous.

At present, there exist asteroid models which can approximate real shape very closely and produce their lightcurves at different viewing and
illumination geometries, also at high phase angles. This motivated us to apply Fourier analysis on asteroid lightcurves. 
Our goal was to derive a set of rules which would help in interpreting lightcurves when deriving rotation periods.
\textit{In other words, we present a useful tool for evaluation of reliabilities of rotation periods based on two parameters.}

\section{Numerical simulations}

We constructed a set of test models using Gaussian random spheres
\citep{Mui+98a} with parameters given by \cite{Mui+98b}.  Those models are based
on real shapes of 14 asteroids and small moons, including both regular (4~Vesta)
and very elongated (1620~Geographos) bodies. While \cite{Mui+98b} yielded for
the maximum degree for spherical harmonics $l=10$, in our models we set $l=7$
which gave us sufficiently good resolution.

To compute light variations for the assumed asteroid models we used
rasterization algorithm commonly used in computer 3D graphics. Models are
represented in a form of vertices (points cloud) and triangles defined on those
vertices. Rasterization process is responsible for constructing a 3D scene based
on input model and projecting facets onto 2D image. Shadowing effects on a
model surface are applied by first composing a scene from light-source point
of view and creating a shadow map. This technique reproduces accurate shadows in
the presence of one light source thus is sufficient for displaying asteroids
models.

Brightness of the model can be computed by summing the pixel values of the
rasterized 2D image. The background pixel values are set to 0 and the model
surface elements have a value computed according to a scattering law.

To mimic reflected light we used a linear combination of Lommel-Seeliger
and Lambert scattering law

\begin{equation}
	S = c_{LS}\frac{\mu \mu_0}{\mu + \mu_0} + c_{L}\mu\mu_0
	\label{eq:scatteringLaw}
\end{equation}
where $\mu$ and $\mu_0$ denotes cosines of the angles between surface normal and
direction to the observer and direction to the Sun respectively. We used
$c_{LS}=0.9$ and $c_{L}=0.1$ coefficients. This law is commonly used in
lightcurve inversion techniques.

To create a lightcurve of the body one needs to specify the light-source,
body and observer positions and the spin axis orientation of the model.  For
a given geometry, model is rotated $\Delta\gamma$ about its spin
axis creating one point on the lightcurve. Process is repeated until a full
rotation is completed.

\begin{figure*}
\centering
\caption{\leftline{Examples of lightcurves and the orientation of their models for the zero rotation phase.}}
\includegraphics[trim={0cm 0cm 5cm 10cm},clip,scale=0.7, angle=270]{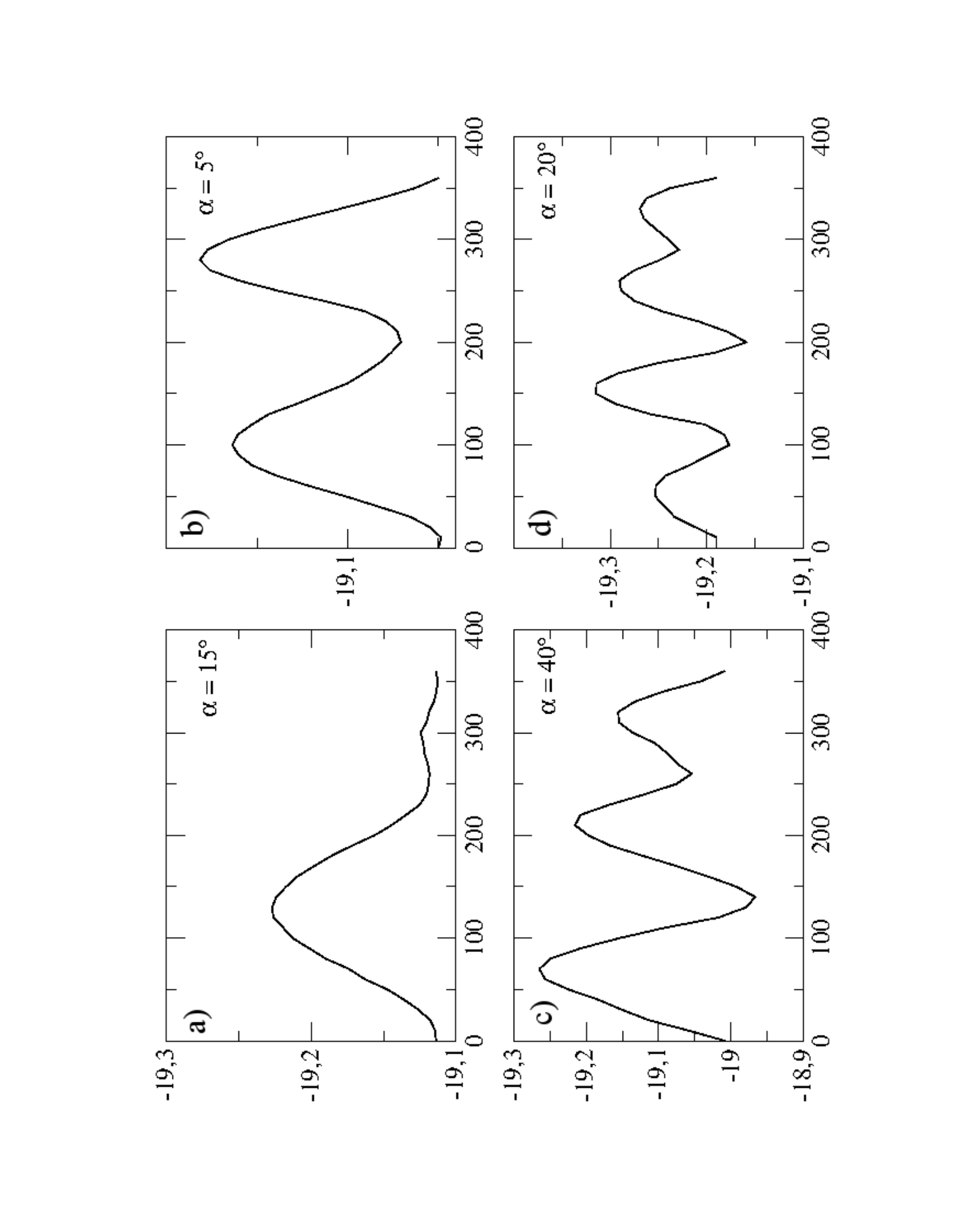}
\includegraphics[trim={0cm 0cm 2cm 10cm},clip,scale=0.6,angle=270]{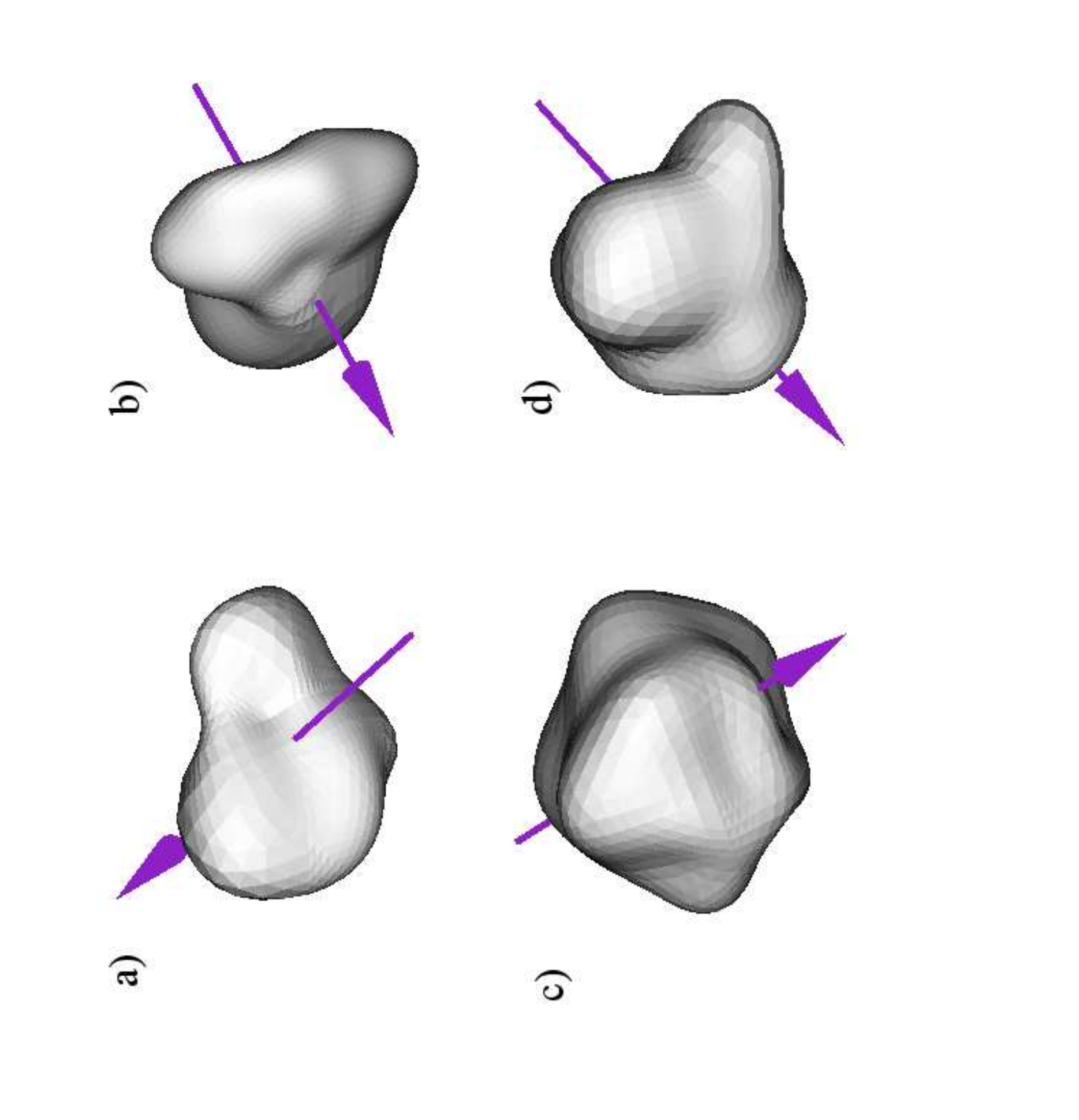}
\label{fig::shapes}
\end{figure*}

The spin axis orientation was given in the reference frame centered at the asteroid,
with the X axis pointing to the observer, and the XY plane defined by the directions of
viewing and illumination. The pole position was measured by the spherical coordinates
of $\lambda$ and $\beta$ ($0 < \lambda < 2\pi$ and $-\rfrac{\pi}{2} < \beta < \rfrac{\pi}{2}$).
To generate random positions of the spin vectors over the whole celestial sphere we used
the following equations \citep{Fel+71}:
\setlength{\mathindent}{2.5cm}
\begin{flalign}
  \centering
  \lambda & = 2\pi u, \label{spin_orient}\\
  \beta & = \arccos(1 - 2v),\notag
\end{flalign}
where $u$ and $v$ are independent uniform random variates on [0, 1).
From the theoretical point of view, the isotropic distribution of pole latitude $\beta$ is envisaged. 
However, \cite{Han+11} has found the clustering of modelled asteroids poles towards the ecliptic poles. They made a statistical analysis of 206
main belt asteroids models and the gap of small latitudes for asteroids with $D < 30$~km was observed. That non-uniform distribution can be explained
by the YORP effect \citep{Rub+00} which is responsible for affecting the spin rates and spin axis orientations of small asteroids.
The other effect can be also taken into account. The deviation from an isotropic distribution of pole latitude can be produced 
by observation and modeling selection effect \citep{Mar+15}.

During our simulations we only considered objects in the principal axis rotation mode.
Also binary and multiple systems were not accounted for. We randomly generated 100 Gaussian
spheres, each of them with 1000 random orientations of the spin axis. For each of them the solar phase angle
was systematically changed from $0\degr$ to $65\degr$ with a step of $5\degr$. After some two
months of computations we built a library of 1~400~000 lightcurves, each of them with 180 points
per rotation cycle. An example of the simulated lightcurve and an asteroid shape used to obtain
it is presented in Fig. \ref{fig::shapes}. On the ISAM webpage\footnote{\url{http://isam.astro.amu.edu.pl}} one can generate animations for 
the rotating models of real asteroids and the resulting lightcurves \citep{Mar+12}.

It is well known that the lightcurve shapes can be approximated by Fourier series \citep{Har+89}
depending on time $t$ and period $P$:
\setlength{\mathindent}{1.5cm}
\begin{flalign} 
  V(t) = & \overline{V} + \sum_{k=1}^n A_k\sin\frac{2\pi k}{P} (t-t_0) \label{Fourier_eq}\\
	      + & B_k\cos\frac{2\pi k}{P}(t-t_0), \notag
\end{flalign}
where $\overline{V}$ is the average brightness, $A_k$ and $B_k$ are Fourier coefficients of the k-th order, 
and $t_0$ is the zero-point time. To parametrize our simulated lightcurves we approximated them with Fourier
series of the 6-th order. For that we used a program described by \cite{Kwi+09}. We have recorded the phase angle
$\alpha$, maximum (peak-to-peak) lightcurve amplitude $A$ and, if $A>0.01$~mag, the amplitudes of each of Fourier
harmonics. Skipping the smallest amplitudes was justified by very irregular shapes of such lightcurves, resulting
from the effect of finite triangular elements approximating surfaces of the models.

\section{Conclusions and Discussion}

At first, we looked at lightcurves with the smallest amplitudes. Table \ref{Tab+amp} presents what 
fraction of such lightcurves were obtained at various phase angles. The $A<0.01$~mag lightcurves, for which
we have not computed Fourier harmonics, are always below 1\%. The fraction of lightcurves with $0.05 < A < 0.1$~mag 
is interesting since, in the presence of noise, they can be interpreted as having no variations at all.
This in turn may suggest that the period is much longer than the time covered during the observations. 
For the Main Belt asteroids, observed usually at $\alpha < 30\degr$, 3~-~8~\% of lightcurves can have amplitudes 
smaller than $0.05$~mag. If the points in our lightcurve are even more scattered, we may overlook realistic light
variations at the level up to $0.1$~mag in 10~-~20~\% of all cases. For near-Earth asteroids, observed often at
$\alpha=30$~-~$50\degr$ the danger of misinterpreting apparently flat lightcurves as an indicator of a long 
period is less probable but still worth considering.

 \begin{figure}
 \centering
 \includegraphics[trim={0 1cm 0 0cm},clip,scale=0.35]{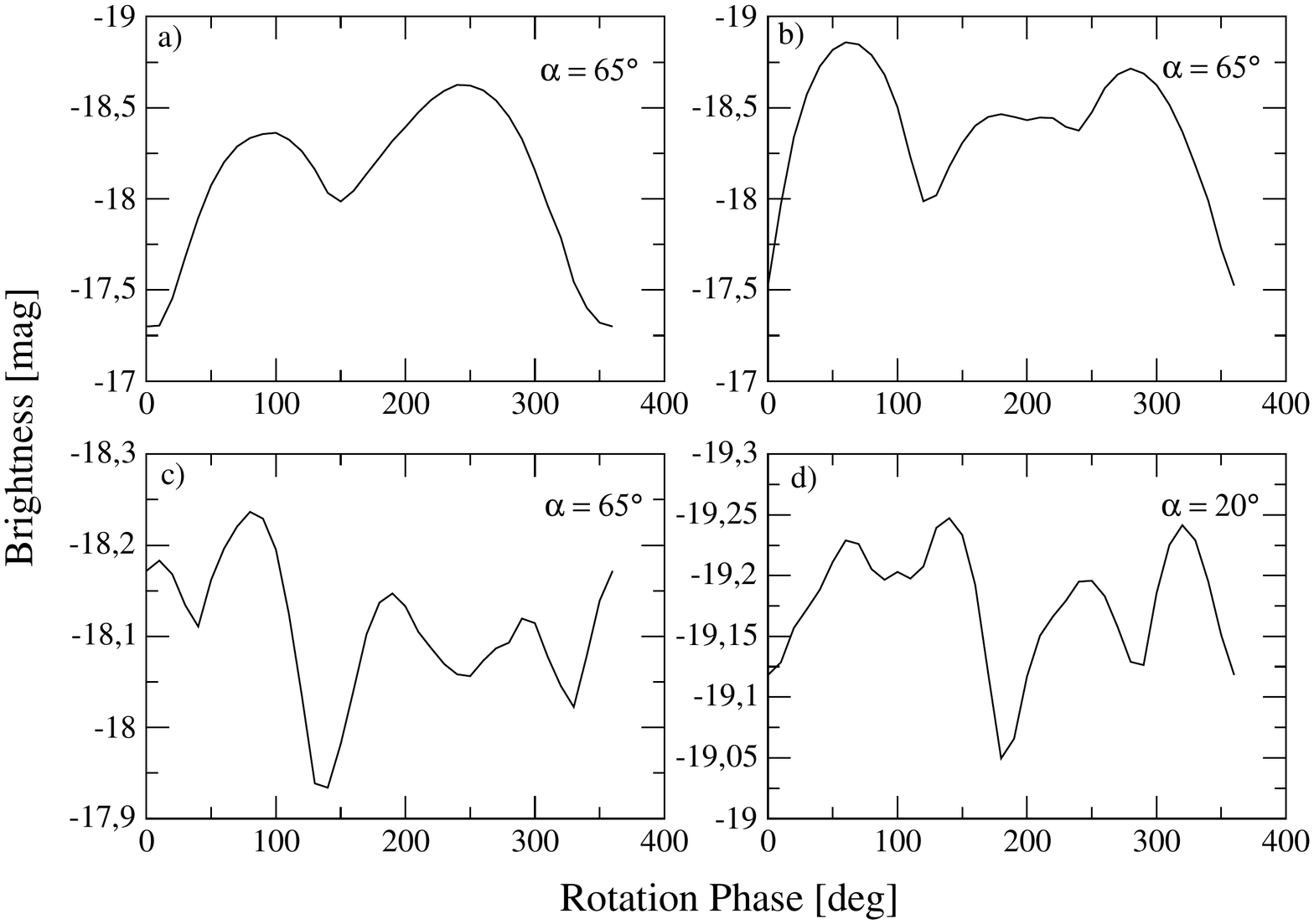}
  \caption[ ]{Example lightcurves obtained at specified phase angles $\alpha$.
\textbf{a)}: $\alpha = 65\degr$, first harmonic dominating, \textbf{b)}: $\alpha = 65\degr$,
third harmonic dominating, \textbf{c)}: $\alpha = 65\degr$, fourth harmonic dominating,
\textbf{d)}: $\alpha = 20\degr$, fourth harmonic dominating.}
 \label{fig:plots}
 \end{figure}

In the next step we tagged each lightcurve according to the number of the Fourier harmonic, whose amplitude was
largest among all six (for convinience we will denote the amplitude of the first harmonic as $A1$, the second as $A2$, etc.
with the maximum, peak-to-peak amplitude marked as $A$). It appeared that $A6$ has never dominated over other harmonics.
$A5$ has been largest only in 223 cases (out of 1~400~000), and only for $\alpha = 35$-$65\degr$. This is statistically
negligible. Because of that we limited our further analysis to the ramaining harmonics: 1st, 2nd, 3rd, and the 4th.

It was obvious to find maximum amplitudes $A$ among lightcurves obtained at $\alpha=65\degr$. The largest amplitude 
observed was $A=1.775$~mag. It was dominated by the second harmonic, and had a typical quasi-sinusoidal shape. For the lightcurves
dominated by remaining three harmonics, 1st, 3rd, and the 4th, the largest amplitudes $A$ were, 1.325, 1.612, and 0.303~mag, respectively
(those lightcurves are presented in Fig.\ref{fig:plots}). It is important to note that the lightcurve dominated by $A1$ (a) seems to have two
maxima and two minima per period. However, secondary minimum is very shallow and while some observers would count it, others would
treat it as a feature on top of the maximum. Because of such subjective measures it is more accurate to refer to the dominating
harmonic instead of the number of pairs of extrema per rotation.

Similar problem can happen when counting maxima on the lightcurve dominated by the third harmonic (Fig. \ref{fig:plots}, b).
However, if we divide the rotation phases into three parts, we can see signs of the three maxima at rotation angles close to
60, 180, and 300$\degr$. Four maxima and minima are better visible on the third lightcurve (Fig. \ref{fig:plots}, c), dominated 
by the fourth harmonic. It is worth noting that our data do not contain any lightcurve with an amplitude $A > 2$~mag. That was the
amplitude of 1620~Geographos, observed in March~1994 at $\alpha=50\degr$ \citep{Mic+94}. However, a closer look at the lightcurve shows
that the deeper of its two minima has a rapid drop, without which its amplitude would be $A=1.8$~mag. This is closer to our simulated 
lightcurve, which at the same phase angle had $A=1.615$~mag. This example shows that while our results are valid in most cases, 
they may not account for every single lightcurve observed.

We refer here to another example of a real asteroid lightcurve. In November-December~2002 asteroid 3155~Lee, observed at $\alpha=20\degr$
displayed a puzzling lightcurve with four pairs of extrema and amplitude $A=0.22$~mag \citep{War+03}. We have several lightcurves of
similar amplitude, which -- at the same phase angle -- are dominated by the fourth harmonic. One of them is shown in Fig. \ref{fig:plots} (d).
However, such lightcurves in our library are rare and account for less than 1\% of all lightcurves obtained at $\alpha < 20\degr$, having
an amplitude of $A < 0.2$~mag.

Also \cite{Mar+15} show another spurious example such as 219~Thusnelda, the main belt object observed from October till December 2014.
The peak-to-peak amplitude was at the level of 0.24~mag and average phase angle $\alpha = 17\degr$.
Previous period for this object was around 29.8~h for a composite lightcurve with one maximum with a ''bump'' before it \citep{Lag+82,Har+92}. 
\cite{Mar+15} obtained the new period for 219~Thusnelda and it is almost two times longer (59.74~h) than previous ones. 
The lightcurve shows two clear maxima with a small ''shelf'' before one of them.

\begin{figure*}
\centering
\caption{Dominating harmonics in asteroid lightcurves.}
\includegraphics[width=0.8\textwidth,angle=270]{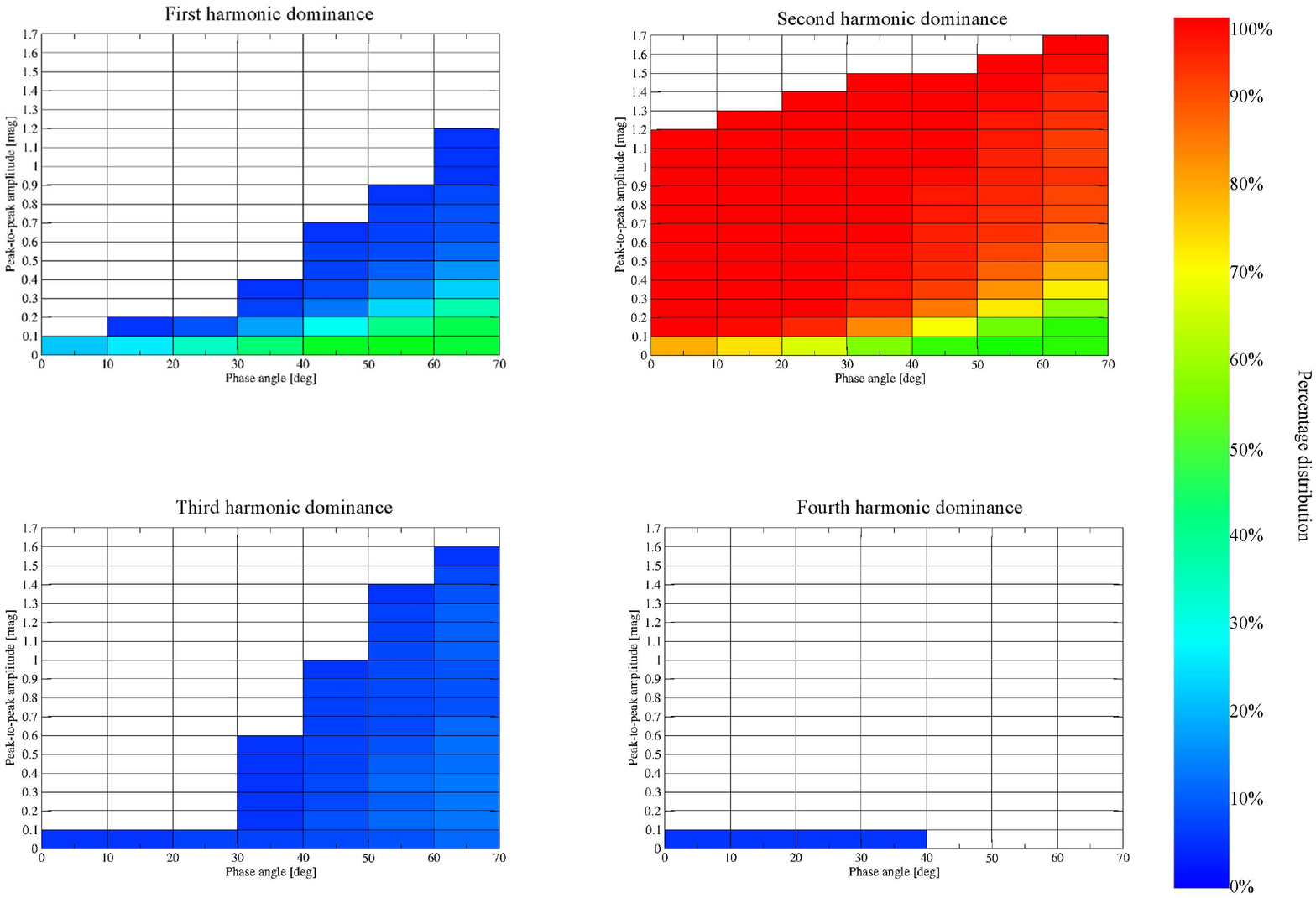}

\medskip
\begin{minipage}{\textwidth} 
{\footnotesize \textbf{Notes.} Fraction of lightcurves (in per cent) in which a given Fourier harmonic is maximal.
White bins refer to lower than 1\% occurance. Both the phase angle and the peak-to-peak amplitude are usually known
when deriving the rotation period.}
\end{minipage}

\label{fig:dist}
\end{figure*}

Instead of matching individual lightcurves, the goal of our simulations was to derive statisticaly significant conclusions about the occurance
of different harmonics in the lightcurves of the specified amplitude $A$ and phase angle $\alpha$. To achieve that we present, on the
$\alpha$-$A$ plane, percentage of lightcurves dominated by the specified harmonic in relation to all lightcurves obtained for the selected
bin (Fig. \ref{fig:dist}, more details in Tab. \ref{tab::first}, \ref{tab::second}, \ref{tab::third}, \ref{tab::fourth}).

We can also draw some general conclusions:
\begin{itemize}
 \item[--] 5th and 6th harmonics never dominate the lightcurve shape, they can only modify the general trends set by lower harmonics,
 \item[--] 4th harmonic responsible for four distinctive maxima and minima dominates in about 1\% of all lightcurves and only at low amplitudes ($A<0.1$~mag),
 \item[--] the dominance of the 3rd harmonic can be observed more often only in cases of near-Earth asteroids (NEAs), which are observed at 
 $\alpha>30\degr$; for the Main Belt asteroids (MBAs) it can be present in small amplitude lightcurves ($A<0.2$~mag),
 \item[--] the 1st harmonic is present quite often in the low amplitude ($A<0.2$~mag) lightcurves of MBAs; for NEAs it can be seen even in high
 amplitude lightcurves ($A<0.7$~mag for $\alpha \simeq 40\degr$, $A<0.9$~mag for $\alpha \simeq 50\degr$),
 \item[--] in practically 100\% of cases the 2nd harmonic dominates the lightcurves of MBAs whose amplitudes $A>0.2$~mag 
 (because they are usually not observable at phase angles $\alpha > 30\degr$).
\end{itemize}

\begin{table}
  \caption{Fraction of lightcurves with low amplitudes.}
  \label{Tab+amp}
  \centering
  \begin{threeparttable}
  \begin{tabular}{rrrr}
  \toprule
  Phase angle [$\degr$]		&$A<0.01$	&$A<0.05$	&$A<0.1$	\\
  \midrule
  0				&0.0046		&0.0785		&0.2161		\\
  5				&0.0042		&0.0764		&0.2087		\\
  10				&0.0026		&0.0735		&0.1964		\\
  15				&0.0010		&0.0645		&0.1800		\\
  20				&0.0002		&0.0553		&0.1647		\\
  25				&0.0001		&0.0418		&0.1468		\\
  30				&0.0000		&0.0294		&0.1222		\\
  35				&0.0000		&0.0188		&0.0994		\\
  40				&0.0000		&0.0115		&0.0770		\\
  45				&0.0000		&0.0070		&0.0583		\\
  50				&0.0000		&0.0037		&0.0407		\\
  55				&0.0000		&0.0022		&0.0267		\\
  60				&0.0000		&0.0014		&0.0183		\\
  65				&0.0000		&0.0010		&0.0128		\\
  \bottomrule
  \end{tabular}
  \begin{tablenotes}[para,flushleft]
  \textbf{Notes.} Columns 2-4 show the fraction of lightcurves (obtained at the specific phase angle)
  whose peak-to-peak amplitudes $A$ (in magnitudes) were smaller than the value given in the header.
  \end{tablenotes}
  \end{threeparttable}
\end{table}

What is worth to notice our analysis is not complete. We do not include objects with non-principal axis or binaries. Our goal was to show in which cases of
determination of rotation period is solid and reliable. The results shown in tables \ref{tab::first}, \ref{tab::second}, \ref{tab::third} can be used 
in individual cases to estimate the probability that the obtained lightcurve is dominated by specific harmonic.
It is a first approximation for those who utilize of lightcurves for further work e.g. reconstruction of the shape and spin axis orientation of asteroids. 
Moreover, one can derive an unique period by adding other observational techniques. For example, \cite{Dur+16} have shown that photometric observations can be merged with thermal data. 
They used sparse photometry contained in the Lowell photometric database and data gained during the Wide-field Infrared Survey Explorer (WISE) mission.
WISE observed asteroids in four filters at wavelengths 3.4, 4.6, 11, and 22 $\mu$m. The first wavelength (W1) corresponds to 
almost 100\% of reflected light. These data typically have about ten measurement points per filter spread over a day or two.
Based only on data from Lowell they get period hidden in many local minima. After adding WISE data the period is more obvious.
Although the results are based on one example, they show that in thermal data reliable information about the rotation period of asteroid can be hidden. 
Going further, using various data types, such as adaptive optics, radar data, stellar occultations we can improve the scientific understanding of small bodies
and abolish ambiguities in determination of rotational period \citep{Mul+17}.

\section*{Acknowledgements}
%
MBB, PB, GD, and AM have recieved funding from the European Union's Horizon 2020 Research and Innovation Programme,
under Grant Agreement no 687378. MBB and TK acknowledge support of the Polish National Science Centre grant N~N203~403739.
PB was supported by the Polish National Science Centre grant N~N203~404139 and partialy supported by grant no. 2014/13/D/ST9/01818 also from the National Science Centre, Poland.
AM was supported by grant no. 2014/13/D/ST9/01818 from the National Science Centre, Poland.




\bibliographystyle{mnras}
\bibliography{rotation} 



%
%


\bsp	

\appendix

\section{Tables - results}


\begin{landscape}
\begin{table}
\centering
\caption{Dominance of the first harmonic in asteroid lightcurves.}
\label{tab::first}
\rowcolors{1}{}{lightgray}
\begin{tabular}{|r|r|r|r|r|r|r|r|r|r|r|r|r|r|r|}
  \hline
A $\symbol{92}$ $\alpha$ & 0.0 - 5\degr & 5 - 10\degr & 10 - 15\degr  & 15 - 20\degr & 20 - 25\degr & 25 - 30\degr & 30 - 35\degr & 35 - 40\degr & 40 - 45\degr & 45 - 50\degr & 50 - 55\degr & 55 - 60\degr & 60 - 65\degr & 65 - 70\degr\\ 
  \hline
  \hline
1.6 - 1.7 & 0.00 &  0.00 &  0.00 &  0.00 &  0.00 &  0.00 &  0.00 &  0.00 &  0.00 &  0.00 & 0.00 & 0.00 & 0.00 & 0.00 \\
1.5 - 1.6 & 0.00 &  0.00 &  0.00 &  0.00 &  0.00 &  0.00 &  0.00 &  0.00 &  0.00 &  0.00 & 0.00 & 0.00 & 0.00 & 0.00 \\
1.4 - 1.5 & 0.00 &  0.00 &  0.00 &  0.00 &  0.00 &  0.00 &  0.00 &  0.00 &  0.00 &  0.00 & 0.00 & 0.00 & 0.00 & 0.00 \\
1.3 - 1.4 & 0.00 &  0.00 &  0.00 &  0.00 &  0.00 &  0.00 &  0.00 &  0.00 &  0.00 &  0.00 & 0.00 & 0.00 & 0.00 & 0.00 \\
\hline                                                                                                       
1.2 - 1.3 & 0.00 &  0.00 &  0.00 &  0.00 &  0.00 &  0.00 &  0.00 &  0.00 &  0.00 &  0.00 & 0.00 & 0.00 & 0.00 & 0.68 \\
1.1 - 1.2 & 0.00 &  0.00 &  0.00 &  0.00 &  0.00 &  0.00 &  0.00 &  0.00 &  0.00 &  0.00 & 0.00 & 0.00 & 0.95 & 1.16 \\
1.0 - 1.1 & 0.00 &  0.00 &  0.00 &  0.00 &  0.00 &  0.00 &  0.00 &  0.00 &  0.00 &  0.00 & 0.00 & 0.27 & 0.42 & 1.20 \\
0.9 - 1.0 & 0.00 &  0.00 &  0.00 &  0.00 &  0.00 &  0.00 &  0.00 &  0.00 &  0.00 &  0.00 & 0.05 & 0.08 & 0.85 & 2.18 \\
\hline                                                                                                       
0.8 - 0.9 & 0.00 &  0.00 &  0.00 &  0.00 &  0.00 &  0.00 &  0.00 &  0.00 &  0.00 &  0.00 & 0.41 & 1.39 & 2.97 & 4.17 \\
0.7 - 0.8 & 0.00 &  0.00 &  0.00 &  0.00 &  0.00 &  0.00 &  0.00 &  0.00 &  0.00 &  0.16 & 1.11 & 2.54 & 3.44 & 4.61 \\
0.6 - 0.7 & 0.00 &  0.00 &  0.00 &  0.00 &  0.00 &  0.00 &  0.00 &  0.00 &  0.41 &  1.15 & 1.98 & 2.69 & 3.50 & 4.46 \\
0.5 - 0.6 & 0.00 &  0.00 &  0.00 &  0.00 &  0.00 &  0.00 &  0.00 &  0.26 &  1.35 &  1.79 & 2.72 & 3.39 & 5.36 & 7.65 \\
\hline                                                                          
0.4 - 0.5 & 0.00 &  0.00 &  0.00 &  0.00 &  0.00 &  0.00 &  0.09 &  0.70 &  1.78 &  2.92 & 4.20 & 6.67 & 8.96 & 13.06 \\
0.3 - 0.4 & 0.00 &  0.00 &  0.00 &  0.00 &  0.00 &  0.07 &  0.50 &  1.17 &  2.27 &  4.39 & 7.38 & 11.42 & 15.80 & 20.14 \\
0.2 - 0.3 & 0.00 &  0.00 &  0.00 &  0.00 &  0.10 &  0.28 &  0.78 &  3.13 &  6.66 &  11.06 & 16.15 & 21.15 & 28.31 & 33.53 \\
0.1 - 0.2 & 0.30 & 0.37 & 0.61 & 1.08 & 3.08 & 5.86 & 9.66 & 15.15 & 20.84 & 27.37 & 32.54 & 37.58 & 40.02 & 41.05 \\
0.0 - 0.1 & 15.82 & 17.79 & 19.49 & 22.58 & 26.24 & 30.65 & 34.97 & 39.60 & 42.98 & 45.10 & 46.29 & 45.18 & 43.88 & 40.39 \\

\hline
\end{tabular}
\end{table}

\begin{table}
\centering
\caption{Dominance of the second harmonic in asteroid lightcurves.}
\label{tab::second}
\rowcolors{1}{}{lightgray}
\begin{tabular}{|r|r|r|r|r|r|r|r|r|r|r|r|r|r|r|}
  \hline
A $\symbol{92}$ $\alpha$ & 0.0 - 5\degr & 5 - 10\degr & 10 - 15\degr  & 15 - 20\degr & 20 - 25\degr & 25 - 30\degr & 30 - 35\degr & 35 - 40\degr & 40 - 45\degr & 45 - 50\degr & 50 - 55\degr & 55 - 60\degr & 60 - 65\degr & 65 - 70\degr \\ 
  \hline
  \hline
1.6 - 1.7 & 0.00 & 0.00 & 0.00 & 0.00 & 0.00 & 0.00 & 0.00 & 0.00 & 0.00 & 0.00 & 0.00 & 0.00 & 100.00 & 100.00 \\
1.5 - 1.6 & 0.00 & 0.00 & 0.00 & 0.00 & 0.00 & 0.00 & 0.00 & 0.00 & 0.00 & 0.00 & 100.00 & 100.00 & 100.00 & 96.80 \\
1.4 - 1.5 & 0.00 & 0.00 & 0.00 & 0.00 & 0.00 & 0.00 & 0.00 & 100.00 & 100.00 & 100.00 & 100.00 & 100.00 & 97.83 & 96.12 \\
1.3 - 1.4 & 0.00 & 0.00 & 0.00 & 0.00 & 100.00 & 100.00 & 100.00 & 100.00 & 100.00 & 100.00 & 100.00 & 97.52 & 97.03 & 95.11 \\
\hline
1.2 - 1.3 & 0.00 & 0.00 & 100.00 & 100.00 & 100.00 & 100.00 & 100.00 & 100.00 & 100.00 & 100.00 & 99.25 & 95.91 & 95.65 & 93.17 \\
1.1 - 1.2 & 0.00 & 100.00 & 100.00 & 100.00 & 100.00 & 100.00 & 100.00 & 100.00 & 100.00 & 100.00 & 99.23 & 96.34 & 93.93 & 94.01 \\
1.0 - 1.1 & 100.00 & 100.00 & 100.00 & 100.00 & 100.00 & 100.00 & 100.00 & 100.00 & 100.00 & 100.00 & 97.78 & 96.21 & 95.63 & 93.87 \\
0.9 - 1.0 & 100.00 & 100.00 & 100.00 & 100.00 & 100.00 & 100.00 & 100.00 & 100.00 & 99.79 & 98.45 & 96.65 & 96.29 & 95.57 & 93.56 \\
\hline
0.8 - 0.9 & 100.00 & 100.00 & 100.00 & 100.00 & 100.00 & 100.00 & 100.00 & 100.00 & 99.58 & 97.26 & 96.54 & 96.18 & 94.23 & 90.72 \\
0.7 - 0.8 & 100.00 & 100.00 & 100.00 & 100.00 & 100.00 & 100.00 & 100.00 & 100.00 & 99.11 & 97.37 & 96.23 & 94.69 & 92.84 & 89.70 \\
0.6 - 0.7 & 100.00 & 100.00 & 100.00 & 100.00 & 100.00 & 100.00 & 100.00 & 99.81 & 97.60 & 96.62 & 95.49 & 93.45 & 91.52  &89.05 \\
0.5 - 0.6 & 100.00 & 100.00 & 100.00 & 100.00 & 100.00 & 100.00 & 99.97 & 98.60 & 97.14 & 96.31 & 94.33 & 92.09 & 88.28 & 84.75 \\
\hline
0.4 - 0.5 & 100.00 & 100.00 & 100.00 & 100.00 & 100.00 & 100.00 & 99.53 & 97.78 & 96.61 & 94.24 & 91.65 & 87.74 & 84.56 & 78.96 \\
0.3 - 0.4 & 100.00 & 100.00 & 100.00 & 100.00 & 100.00 & 99.79 & 98.28 & 97.42 & 95.59 & 91.84 & 88.02 & 82.18 & 76.67 & 71.38 \\
0.2 - 0.3 & 100.00 & 100.00 & 100.00 & 100.00 & 99.89 & 99.52 & 98.69 & 95.48 & 90.69 & 85.60 & 79.29 & 72.40 & 64.38 & 58.86 \\
0.1 - 0.2 & 99.69 & 99.62 & 99.38 & 98.91 & 96.91 & 93.90 & 89.26 & 82.40 & 75.46 & 68.17 & 61.80 & 55.37 & 52.37 & 50.30 \\
0.0 - 0.1 & 82.49 & 80.23 & 78.23 & 74.75 & 71.06 & 66.62 & 62.08 & 57.49 & 53.88 & 51.84 & 49.76 & 49.36 & 50.58 & 53.10 \\
   \hline
\end{tabular}
\end{table}
\end{landscape}

\begin{landscape}
\begin{table}
\centering
\caption{Dominance of the third harmonic in asteroid lightcurves.}
\label{tab::third}
\rowcolors{1}{}{lightgray}
\begin{tabular}{|r|r|r|r|r|r|r|r|r|r|r|r|r|r|r|}
  \hline
A $\symbol{92}$ $\alpha$ & 0.0 - 5\degr & 5 - 10\degr & 10 - 15\degr  & 15 - 20\degr & 20 - 25\degr & 25 - 30\degr & 30 - 35\degr & 35 - 40\degr & 40 - 45\degr & 45 - 50\degr & 50 - 55\degr & 55 - 60\degr & 60 - 65\degr & 65 - 70\degr\\ 
  \hline
  \hline
1.6 - 1.7 &  0.00 &   0.00 &   0.00 &   0.00 &   0.00 &   0.00 &   0.00 &   0.00 &   0.00 &   0.00 &   0.00 &   0.00 &   0.00 &   0.00 \\
1.5 - 1.6 &  0.00 &   0.00 &   0.00 &   0.00 &   0.00 &   0.00 &   0.00 &   0.00 &   0.00 &   0.00 &   0.00 &   0.00 &   0.00 &   3.19 \\
1.4 - 1.5 &  0.00 &   0.00 &   0.00 &   0.00 &   0.00 &   0.00 &   0.00 &   0.00 &   0.00 &   0.00 &   0.00 &   0.00 &   2.16 &   3.87 \\
1.3 - 1.4 &  0.00 &   0.00 &   0.00 &   0.00 &   0.00 &   0.00 &   0.00 &   0.00 &   0.00 &   0.00 &   0.00 &   2.48 &   2.97 &   4.89 \\
\hline                                                                                                                       
1.2 - 1.3 &  0.00 &   0.00 &   0.00 &   0.00 &   0.00 &   0.00 &   0.00 &   0.00 &   0.00 &   0.00 &   0.74 &   4.08 &   4.34 &   6.14 \\
1.1 - 1.2 &  0.00 &   0.00 &   0.00 &   0.00 &   0.00 &   0.00 &   0.00 &   0.00 &   0.00 &   0.00 &   0.76 &   3.65 &   5.11 &   4.82 \\
1.0 - 1.1 &  0.00 &   0.00 &   0.00 &   0.00 &   0.00 &   0.00 &   0.00 &   0.00 &   0.00 &   0.00 &   2.22 &   3.51 &   3.94 &   4.92 \\
0.9 - 1.0 &  0.00 &   0.00 &   0.00 &   0.00 &   0.00 &   0.00 &   0.00 &   0.00 &   0.20 &   1.55 &   3.29 &   3.62 &   3.57 &   4.24 \\
\hline                                                                                                                       
0.8 - 0.9 &  0.00 &   0.00 &   0.00 &   0.00 &   0.00 &   0.00 &   0.00 &   0.00 &   0.42 &   2.74 &   3.04 &   2.41 &   2.79 &   5.10 \\
0.7 - 0.8 &  0.00 &   0.00 &   0.00 &   0.00 &   0.00 &   0.00 &   0.00 &   0.00 &   0.88 &   2.47 &   2.66 &   2.76 &   3.71 &   5.69 \\
0.6 - 0.7 &  0.00 &   0.00 &   0.00 &   0.00 &   0.00 &   0.00 &   0.00 &   0.18 &   1.98 &   2.23 &   2.52 &   3.85 &   4.97 &   6.48 \\
0.5 - 0.6 &  0.00 &   0.00 &   0.00 &   0.00 &   0.00 &   0.00 &   0.02 &   1.13 &   1.50 &   1.89 &   2.94 &   4.50 &   6.36 &   7.59 \\
\hline                                                                                                                       
0.4 - 0.5 &  0.00 &   0.00 &   0.00 &   0.00 &   0.00 &   0.00 &   0.38 &   1.51 &   1.60 &   2.83 &   4.14 &   5.58 &   6.47 &   7.97 \\
0.3 - 0.4 &  0.00 &   0.00 &   0.00 &   0.00 &   0.00 &   0.13 &   1.21 &   1.41 &   2.12 &   3.76 &   4.60 &   6.39 &   7.52 &   8.47 \\
0.2 - 0.3 &  0.00 &   0.00 &   0.00 &   0.00 &   0.00 &   0.19 &   0.52 &   1.38 &   2.63 &   3.32 &   4.54 &   6.44 &   7.29 &   7.59 \\
0.1 - 0.2 &  0.00 &   0.00 &   0.00 &   0.00 &   0.00 &   0.15 &   0.87 &   2.28 &   3.59 &   4.30 &   5.37 &   6.72 &   7.11 &   8.15 \\
0.0 - 0.1 &  0.69 &   0.91 &   1.16 &   1.50 &   1.61 &   1.86 &   2.21 &   2.39 &   2.50 &   2.62 &   3.53 &   5.05 &   5.30 &   6.07 \\

   \hline
\end{tabular}
\end{table}
%
 \begin{table}
 \centering
 \caption{Dominance of the fourth harmonic in asteroid lightcurves.}
 \label{tab::fourth}
 \rowcolors{1}{}{lightgray}
 \begin{tabular}{|r|r|r|r|r|r|r|r|r|r|r|r|r|r|r|}
   \hline
 A $\symbol{92}$ $\alpha$ & 0.0 - 5\degr & 5 - 10\degr & 10 - 15\degr  & 15 - 20\degr & 20 - 25\degr & 25 - 30\degr & 30 - 35\degr & 35 - 40\degr & 40 - 45\degr & 45 - 50\degr & 50 - 55\degr & 55 - 60\degr & 60 - 65\degr & 65 - 70\degr\\ 
   \hline
   \hline
1.6 - 1.7 &  0.00 &   0.00 &   0.00 &   0.00 &   0.00 &   0.00 &   0.00 &   0.00 &   0.00 &   0.00 &   0.00 &   0.00 &   0.00 &   0.00 \\
1.5 - 1.6 &  0.00 &   0.00 &   0.00 &   0.00 &   0.00 &   0.00 &   0.00 &   0.00 &   0.00 &   0.00 &   0.00 &   0.00 &   0.00 &   0.00 \\
1.4 - 1.5 &  0.00 &   0.00 &   0.00 &   0.00 &   0.00 &   0.00 &   0.00 &   0.00 &   0.00 &   0.00 &   0.00 &   0.00 &   0.00 &   0.00 \\
1.3 - 1.4 &  0.00 &   0.00 &   0.00 &   0.00 &   0.00 &   0.00 &   0.00 &   0.00 &   0.00 &   0.00 &   0.00 &   0.00 &   0.00 &   0.00 \\
\hline
1.2 - 1.3 &  0.00 &   0.00 &   0.00 &   0.00 &   0.00 &   0.00 &   0.00 &   0.00 &   0.00 &   0.00 &   0.00 &   0.00 &   0.00 &   0.00 \\
1.1 - 1.2 &  0.00 &   0.00 &   0.00 &   0.00 &   0.00 &   0.00 &   0.00 &   0.00 &   0.00 &   0.00 &   0.00 &   0.00 &   0.00 &   0.00 \\
1.0 - 1.1 &  0.00 &   0.00 &   0.00 &   0.00 &   0.00 &   0.00 &   0.00 &   0.00 &   0.00 &   0.00 &   0.00 &   0.00 &   0.00 &   0.00 \\
0.9 - 1.0 &  0.00 &   0.00 &   0.00 &   0.00 &   0.00 &   0.00 &   0.00 &   0.00 &   0.00 &   0.00 &   0.00 &   0.00 &   0.00 &   0.00 \\
\hline                                                                                                                       
0.8 - 0.9 &  0.00 &   0.00 &   0.00 &   0.00 &   0.00 &   0.00 &   0.00 &   0.00 &   0.00 &   0.00 &   0.00 &   0.00 &   0.00 &   0.00 \\
0.7 - 0.8 &  0.00 &   0.00 &   0.00 &   0.00 &   0.00 &   0.00 &   0.00 &   0.00 &   0.00 &   0.00 &   0.00 &   0.00 &   0.00 &   0.00 \\
0.6 - 0.7 &  0.00 &   0.00 &   0.00 &   0.00 &   0.00 &   0.00 &   0.00 &   0.00 &   0.00 &   0.00 &   0.00 &   0.00 &   0.00 &   0.00 \\
0.5 - 0.6 &  0.00 &   0.00 &   0.00 &   0.00 &   0.00 &   0.00 &   0.00 &   0.00 &   0.00 &   0.00 &   0.00 &   0.00 &   0.00 &   0.00 \\
\hline                                                                                                                       
0.4 - 0.5 &  0.00 &   0.00 &   0.00 &   0.00 &   0.00 &   0.00 &   0.00 &   0.00 &   0.00 &   0.00 &   0.00 &   0.00 &   0.00 &   0.00 \\
0.3 - 0.4 &  0.00 &   0.00 &   0.00 &   0.00 &   0.00 &   0.00 &   0.00 &   0.00 &   0.00 &   0.00 &   0.00 &   0.00 &   0.00 &   0.00 \\
0.2 - 0.3 &  0.00 &   0.00 &   0.00 &   0.00 &   0.00 &   0.00 &   0.00 &   0.00 &   0.00 &   0.00 &   0.00 &   0.00 &   0.01 &   0.01 \\
0.1 - 0.2 &  0.00 &   0.00 &   0.00 &   0.00 &   0.00 &   0.08 &   0.20 &   0.17 &   0.10 &   0.11 &   0.25 &   0.19 &   0.36 &   0.36 \\
0.0 - 0.1 &  0.99 &   1.06 &   1.11 &   1.16 &   1.07 &   0.86 &   0.73 &   0.47 &   0.53 &   0.32 &   0.23 &   0.18 &   0.03 &   0.13 \\

    \hline
 \end{tabular}
 \end{table}
\end{landscape}
\label{lastpage}
\end{document}